\documentclass{bmcart}

\usepackage{cite}
\usepackage{todonotes}
\usepackage{multirow}
\usepackage{graphicx}
\usepackage{tikz,pgfplots}
\usepackage{pgfplotstable}
\usepackage{subfigure}
\usepackage{caption}
\usepackage{graphicx,pgfplots}
\usepackage{tikz,pgfplots}
\usepackage{pgfplotstable}
\usepackage{comment}
\usepackage{url}
\usepackage[english]{babel}
\usepackage[utf8x]{inputenc}
\usepackage{hyperref}

\startlocaldefs
\endlocaldefs

\begin{document}

\begin{frontmatter}

\begin{fmbox}
\dochead{Research}


\title{Exposing Influence Campaigns in the Age of LLMs: A Behavioral-Based AI Approach to Detecting State-Sponsored Trolls}


\author[
   addressref={aff1,aff2},                   
corref={aff1},                       
   email={fatima.ezzeddine@supsi.ch}   
]{\inits{FE}\fnm{Fatima} \snm{Ezzeddine}}
\author[
   addressref= {aff1},
   email={omran.ayoub@supsi.ch}
]{\inits{OA}\fnm{Omran} \snm{Ayoub}}
\author[
  addressref= {aff1},
  email={silvia.giordano@supsi.ch}
]{\inits{SG}\fnm{Silvia} \snm{Giordano}}
\author[
  addressref= {aff1},
  email={gianluca.nogara@supsi.ch}
]{\inits{GN}\fnm{Gianluca} \snm{Nogara}}
\author[
  addressref= {aff2},
  email={ihab.sbeity@gmail.com}
]{\inits{IS}\fnm{Ihab} \snm{Sbeity}}
\author[
  addressref= {aff3},
  email={emiliofe@usc.edu}
]{\inits{EF}\fnm{Emilio} \snm{Ferrara}}
\author[
  addressref= {aff1,aff3},
  email={lluceri@isi.edu}
]{\inits{LL}\fnm{Luca} \snm{Luceri}}

\address[id=aff1]{%
  \orgname{University of Applied Sciences and Arts of Southern Switzerland, Department of Innovative Technologies},
  \city{Lugano},
  \cny{Switzerland}
}

\address[id=aff2]{
  \orgname{Lebanese University, Faculty of Science, Department of Applied Mathematics}, 
  \city{Beirut},                              
  \cny{Lebanon}                                    
}

\address[id=aff3]{%
  \orgname{Information Sciences Institute, Viterbi School of Engineering, University of Southern California},
  \city{Marina del Rey, CA},
  \cny{USA}
}



\end{fmbox}


\begin{abstractbox}

\begin{abstract} 
The detection of state-sponsored trolls operating in influence campaigns on social media is a critical and unsolved challenge for the research community, which has significant implications beyond the online realm. To address this challenge, we propose a new AI-based solution that identifies troll accounts solely through behavioral cues associated with their sequences of sharing activity, encompassing both their actions and the feedback they receive from others. 
Our approach does not incorporate any textual content shared
and consists of two steps: First, we leverage an LSTM-based classifier to determine whether account sequences belong to a state-sponsored troll or an organic, legitimate user. Second, we employ the classified sequences to calculate a metric named the ``Troll Score'', quantifying the degree to which an account exhibits troll-like behavior.
To assess the effectiveness of our method, we examine its performance in the context of the 2016 Russian interference campaign during the U.S. Presidential election. Our experiments yield compelling results, demonstrating that our approach can identify account sequences with an AUC close to 99\% and accurately differentiate between Russian trolls and organic users with an AUC of 91\%. Notably, our behavioral-based approach holds a significant advantage in the ever-evolving landscape, where textual and linguistic properties can be easily mimicked by Large Language Models (LLMs): In contrast to existing language-based techniques, it relies on more challenging-to-replicate behavioral cues, ensuring greater resilience in identifying influence campaigns, especially given the potential increase in the usage of LLMs for generating inauthentic content.
Finally, we assessed the generalizability of our solution to various entities driving different information operations and found promising results that will guide future research.

\end{abstract}




\end{abstractbox}

\end{frontmatter}




\section{Introduction}

Social Media Networks (SMNs) are a crucial constituent of societies, providing a primary platform for individuals to engage in social and political discourse, as well as to disseminate critical messages and promote propaganda. SMNs have undergone a significant transformation, evolving from a simple aggregation medium to a complex ecosystem where the line between offline and online realms is often blurred \cite{luceri2021social}. Recent studies have shown that the impact of discussions on SMNs extends beyond the online platform and can have a significant effect on societies, such as undermining the integrity of political elections and public health \cite{aro2016cyberspace, zollo2015emotional, pariser2011filter, luceri2021down, pierri2022online}.

In this context, the accuracy, confidentiality, and authenticity of shared content are crucial elements for safe communication and, therefore, the well-being of societies. However, SMNs have experienced a shortage of these elements, as their growth has led to an increase in deceptive and fraudulent accounts that intentionally damage the credibility of online discussions \cite{ferrara2015manipulation,luceri2019red}. The activity of these accounts often results in online harms that threaten the honesty and ethics of conversations, such as the propagation of hate speech, incitement of violence, and dissemination of misleading and controversial content. This has been observed in recent debates concerning the Ukraine-Russia conflict \cite{pierri2023propaganda}, Covid-19 pandemic \cite{ferrara2020misinformation, diseases2020covid, hu2020covid, nogara2022disinformation, pierri2022one}, as well as the rise of conspiracy theories \cite{wang2022identifying, suresh2023tracking, phadke2022pathways}. These fraudulent accounts represent a significant threat to healthy online conversations, whose activity has the potential to exacerbate societal divisions and affect the sovereignty of elections \cite{allem2017cigarette, del2016echo, matakos2017measuring, vosoughi2018spread, metaxas2012social,gatta2023interconnected}.

In the political sphere, Russian meddling in the 2016 U.S. Presidential election represents the most prominent case of deceptive online interference campaign \cite{carroll2017st,popken2018twitter}. The Mueller report \cite{mueller2019mueller} suggests that Russia engaged in extensive attacks on the U.S. election system to manipulate the outcome of the 2016 voting event. The ``sweeping and systematic'' interference allegedly used bots (i.e., automated accounts) and trolls (i.e., state-sponsored human operators) to spread politically biased and false information \cite{lopez2018so}. 
In the aftermath of the election, the U.S. Congress released a list of 2,752 Twitter accounts associated with Russia's ``Internet Research Agency" (IRA), known as Russian trolls. As a result, significant research efforts were launched to identify fraudulent accounts and deceptive activity on several SMNs. Among these platforms, Twitter has been continuously working to eliminate malicious entities involved in information operations across different countries\cite{gadde2020additional,alizadeh2020content,nwala2022general} and different geopolitical events \cite{pierri2022does,luceri2019evolution}. While there are several proven techniques for uncovering bot accounts \cite{kudugunta2018deep, ferrara2016rise, mazza2019rtbust, chavoshi2016debot, abou2019graph,cresci2017social,ferrara2022twitter}, the detection of troll accounts is currently an unsolved issue for the research community, due to several factors tied with the human character of trolls \cite{zannettou2019let}.
Note that throughout this manuscript, our definition of \emph{troll} is limited to state-sponsored human actors who have a political agenda and operate in coordinated influence campaigns, disregarding thus other hateful and harassing online activities tied with Internet-mediated trolling behavior.

Recent efforts have devised approaches for identifying trolls by leveraging linguistic cues and profile meta-data \cite{im2020still, badawy2019characterizing,alhazbi2020behavior,saeed2021trollmagnifier,mazza2022investigating}. Although these approaches have shown promising results, they suffer from certain limitations. Some of these methods are language-dependent, focusing solely on specific spoken languages associated with the trolls under investigation \cite{addawood2019linguistic, jachim2020trollhunter}. Others are constrained to a single SMN, relying on profile metadata and platform-specific information. Furthermore, the ease of imitating language and linguistic cues has increased with the emergence of Large Language Models (LLMs), such as ChatGPT and similar technologies. As we look ahead, our ability to detect influence operations based solely on linguistic cues may be hindered by the growing reliance on LLMs for such operations \cite{yang2023anatomy,ferrara2023social}. These significant limitations have prompted research efforts to develop language- and content-agnostic approaches, as demonstrated in the work of Luceri et al. \cite{luceri2020detecting}. This approach distinguishes troll accounts by uncovering behavioral incentives from their observed activities using an Inverse Reinforcement Learning (IRL) framework. Given that mimicking behaviors and incentives is notably more challenging than imitating language, incorporating behavioral cues either in addition to or as an alternative to purely linguistic-based methods emerges as a promising strategy in an uncertain future, particularly when the cost of generating inauthentic, yet credible, content appears to be exceptionally low \cite{menczer2023addressing,mitrovic2023chatgpt}.

In this work, we advance along this research line and propose a novel approach to identify state-sponsored troll activity solely based on behavioral cues linked to accounts' sharing activities on Twitter. Specifically, we consider online activities regardless of the content shared, the language used, and the linked metadata to classify accounts as trolls or organic, legitimate users (from now on, simply \emph{users}). Our approach aims to capture cues of behavior that differentiate trolls from users by analyzing their interactions and responses to feedback. For this purpose, we consider both the actions performed by an account, namely \textit{active online activities}, and the feedback received by others, namely \textit{passive online activities}, e.g., received replies and retweets. We refer to the sequence of active and passive activities as a \emph{trajectory}, in accordance with \cite{luceri2020detecting}. We demonstrate the validity of our approach by detecting Russian trolls involved in the interference of the 2016 U.S. Presidential election. We also evaluate whether the proposed approach can be effectively used to identify various entities involved in diverse Twitter information operations during the 2020 U.S. Presidential election.


\textbf{Contributions of this work.}
The core contributions of this work are summarized as follows:
\begin{itemize}
    \item We propose a novel approach based on Long Short-Term Memory (LSTM) for classifying accounts' trajectories. 
    Our approach correctly identifies trolls' and users' trajectories with an AUC and an F1-score of about 99\%. 
    
    \item Leveraging the classified trajectories, we introduce a metric, namely the \emph{Troll Score}, that enables us to quantitatively assess the extent to which an account exhibits behavior akin to that of a state-sponsored troll. We propose a \emph{Troll Score}-based classifier that can effectively detect troll accounts with remarkable accuracy, achieving an AUC of about 91\% (F1-score $\sim$90\%). Our approach outperforms existing behavioral-based methods and approaches the classification performance of existing linguistic solutions, all while not requiring access to the content of shared messages. This feature enhances its robustness, especially given the possibility of increased usage of LLMs for influence operations.


    \item By analyzing the active and passive activities in which accounts engage, we uncovered three distinct, naturally emerging \textit{behavioral} clusters where trolls intermingle with user accounts. This finding confirms the difficulty of differentiating these two account classes when their trajectories are not considered.

    \item 
    We demonstrate the capability of our approach to generalize and accurately identify diverse actors responsible for driving information operations. The results reveal that our methodology achieves an AUC of 80\% (F1-score $\sim$82\%) in detecting the drivers of different campaigns, indicating promising results for its applicability across countries, languages, and various malicious entities.
\end{itemize}

\section{Related Work}\label{relatedwork}

In this Section, we survey research on the automated detection of malicious accounts operated by trolls, with a focus on the troll farm connected to the IRA \cite{frommer2019twitter}. Some of these efforts have proposed linguistic approaches that rely on the content posted by trolls to identify and detect them. For instance, \cite{addawood2019linguistic} presented a theory-driven linguistic study of Russian trolls' language and demonstrated how deceptive linguistic signals can contribute to accurate troll identification. Similarly, \cite{jachim2020trollhunter} proposed an automated reasoning mechanism for hunting trolls on Twitter during the COVID-19 pandemic, which leverages a unique linguistic analysis based on adversarial machine learning and ambient tactical deception. In \cite{weller2019identifying}, the authors proposed a deep learning solution for troll detection on Reddit and analyzed the shared content using natural language processing techniques. Other works have considered fusing users' metadata and linguistic features, such as \cite{im2020still}, which used profile description, stop word usage, language distribution, and bag of words features for detecting Russian trolls. Other approaches have relied on multimedia analysis, combining text, audio, and video analysis to detect improper material or behavior \cite{vanhove2013towards, valldor2018firearm}. For instance, \cite{vanhove2013towards} designed a platform for monitoring social media networks with the aim of automatically tracking malicious content by analyzing images, videos, and other media. In \cite{valldor2018firearm}, the authors attempted to capture disinformation and trolls based on the existence of a firearm in images using the Object Detection API. A limitation of these works is their reliance on the content posted by accounts and on the extraction of linguistic features for troll identification. In contrast, our approach solely relies on the online behavior of accounts, specifically, the temporal sequence of online activities performed by a user. This presents an advantage over previous works, as it is independent of the language used or content shared and has, therefore, the potential to generalize to influence campaigns originating from diverse countries and be resilient to the use of LLMs for generating inauthentic content.

Previous studies have proposed sequence analysis approaches for identifying malicious accounts. For example, Kim et al. \cite{kim2019analysing} used text and time as features to categorize trolls into subgroups based on the temporal and semantic similarity of their shared content. Luceri et al. \cite{luceri2020detecting} proposed a solution that only relies on the sequence of users' activity on online platforms to capture the incentives that the two classes of accounts (trolls vs. users) respond to. They detect troll accounts with a supervised learning approach fed by the incentives estimated via Inverse Reinforcement Learning (IRL). In \cite{wang2016unsupervised}, the authors proposed a model based on users' sequence of online actions to identify clusters of accounts with similar behavior. However, this approach was found to be ineffective in detecting Russian trolls, as reported in \cite{luceri2020detecting}.
Similarly to these approaches, we propose a language- and content-agnostic method for identifying trolls based only on the sharing activities performed by the accounts on Twitter. We utilize deep learning, specifically LSTM, to classify the sequence of activities as belonging to either troll accounts or organic users. We leverage the classified sequences to quantify the extent to which an account behaves like a troll, a feature not available in earlier methods. 



\section{Problem Formulation and Trajectory Definition}\label{problemtrajectory}

This section outlines the objectives of the proposed framework and elucidates the features, variables, and learning tasks integral to it. While existing methods for identifying troll activity in SMNs rely on linguistic and metadata features, our approach strives to be language- and content-agnostic. To achieve this, we rely only on behavioral cues and do not incorporate any textual or media content shared by the accounts, nor do we use their profile metadata. Consequently, our approach holds the potential for application across various SMNs and is robust against the increasing use of LLMs and their potential role in influence campaigns \cite{yang2023anatomy, menczer2023addressing,ferrara2023social}. 

To extract the unique online behaviors of trolls and organic users on Twitter, we extract the accounts' sequences of online activities. We consider their \emph{active online activities}, including generating an original post (i.e., \emph{tweet}), re-sharing another post (i.e., \emph{retweet}), commenting an existing post (i.e., \emph{reply}),  or mentioning another user in a post (i.e., \emph{mention}). In addition, we also propose to consider the feedback the accounts receive from other accounts, namely \emph{passive online activities}, such as receiving a retweet (i.e., \emph{being retweeted}), a comment (i.e., \emph{being replied to}), or a mention (i.e., \emph{being mentioned in a post}).
By considering both the actions performed by the accounts and the feedback received, we aim to capture the distinct motivations driving trolls' activity, which may differ from those of organic users \cite{luceri2020detecting}. The rationale is that trolls might be motivated to pursue their agenda regardless of the feedback received from others, while organic users may be influenced by the level of endorsement they receive from the online community. For example, users might be more motivated to generate a new tweet when their content is re-shared by others, which is also viewed as a form of social endorsement \cite{metaxas2015retweets,stella2018bots}, or when they receive positive comments. 

To formalize this approach, we model the SMN Twitter as a Markov Decision Process (MDP). Similarly to Luceri et al. \cite{luceri2020detecting}, we represent Twitter as an environment constituted of multiple agents (i.e., Twitter accounts) that can perform a set of actions
(i.e., \emph{active online activities}) and receive feedback from the environment (i.e., \emph{passive online activities}). Consistently with the IRL formulation in \cite{luceri2020detecting}, we refer to the latter as
\emph{states}, as they represent the response of the Twitter environment, whereas we refer to the former as \emph{actions}, as they indicate the \emph{active online activities} that an account can perform on Twitter. 

We consider four \emph{actions} that can be performed by Twitter accounts: 
\begin{itemize}
  \item Original tweet (tw): to generate original content;
  \item Retweet (rt): to re-share content generated by others;
  \item Interact with others (in): to interact with other users via replies or mentions;
  \item No Action (no): to keep silent, i.e., the account does not perform any action.
\end{itemize}

For what pertains to \emph{states}, we consider three possible feedback that Twitter accounts can receive: 
\begin{itemize}
\item Retweeted (RT): an original tweet generated by the account is re-shared; 
\item Interacted with (IN): the account is involved by others via replies (i.e., comments to a tweet generated by the account) or mentions;
\item No Interaction (NO): no feedback is received by the account.
\end{itemize}

Every account can move from one state to another when performing an action, and we refer to such a transition as a \emph{state-action pair}. Note that an account can be only in one of the above-mentioned states and can perform only one action in any given state. By considering the accounts' timeline, we construct a \emph{sequence} of state-action pairs that reconstruct the (observed) history of the account on Twitter. Overall, there exist only 11 possible combinations of state-action pairs that form a sequence of an account--- the state-action pair \emph(NO, no) is not considered as it does not describe any observable activity. It is also important to note that if an account does not react to the environment feedback, e.g., it does not perform \emph{tw}, \emph{rt} or \emph{in}, it will be considered as silent, i.e., doing no action (\emph{no}). Similarly, if an account keeps performing actions while not receiving any feedback, it will persist in the state \emph{NO}.

By extracting and sorting the state-action pairs of every account in chronological order, we create a sequence of online activities describing the behavior of the observed trolls and users. Accordingly, we define the problem of classifying account sequences as a binary classification task with two classes: \emph{troll} (positive class) and \emph{user} (negative class). Section \ref{method} details how we first classify these formed sequences as belonging to either trolls or users and then how we use these classified sequences to identify troll accounts. While our focus is on Twitter, this approach may be replicated on other SMNs like Facebook, which offer similar sharing activities.

\section{Data}\label{data}

The dataset used to evaluate our approach
consists of tweets generated by user accounts and by accounts identified as trolls involved in the 2016 U.S. election discussion over Twitter. Specifically, we consider the dataset described in \cite{addawood2019linguistic, badawy2019characterizing} collected
by utilizing a set of keywords (detailed in \cite{bessi2016social}) related to the 2016 U.S. election, which encompasses 13.5M tweets generated by 2.4M accounts.

In the dataset used for this study, we only included accounts that had generated a minimum of ten active online activities and ten passive online activities. This decision was based on the findings of a previous study \cite{luceri2020detecting}, which demonstrated that using fewer than ten active and passive online activities had a negative impact on classification accuracy.
Among these accounts, we identified 342 troll accounts---selected from a list of 2,752 Twitter accounts ascertained as Russian trolls by the U.S. Congress and that was publicly released during the investigation of Russian involvement in the U.S. 2016 Presidential election; and 1,981 user accounts which generated 246k and 1.2M tweets, respectively. 

\begin{figure*}[t]
    \centering
    \includegraphics[width=360pt]{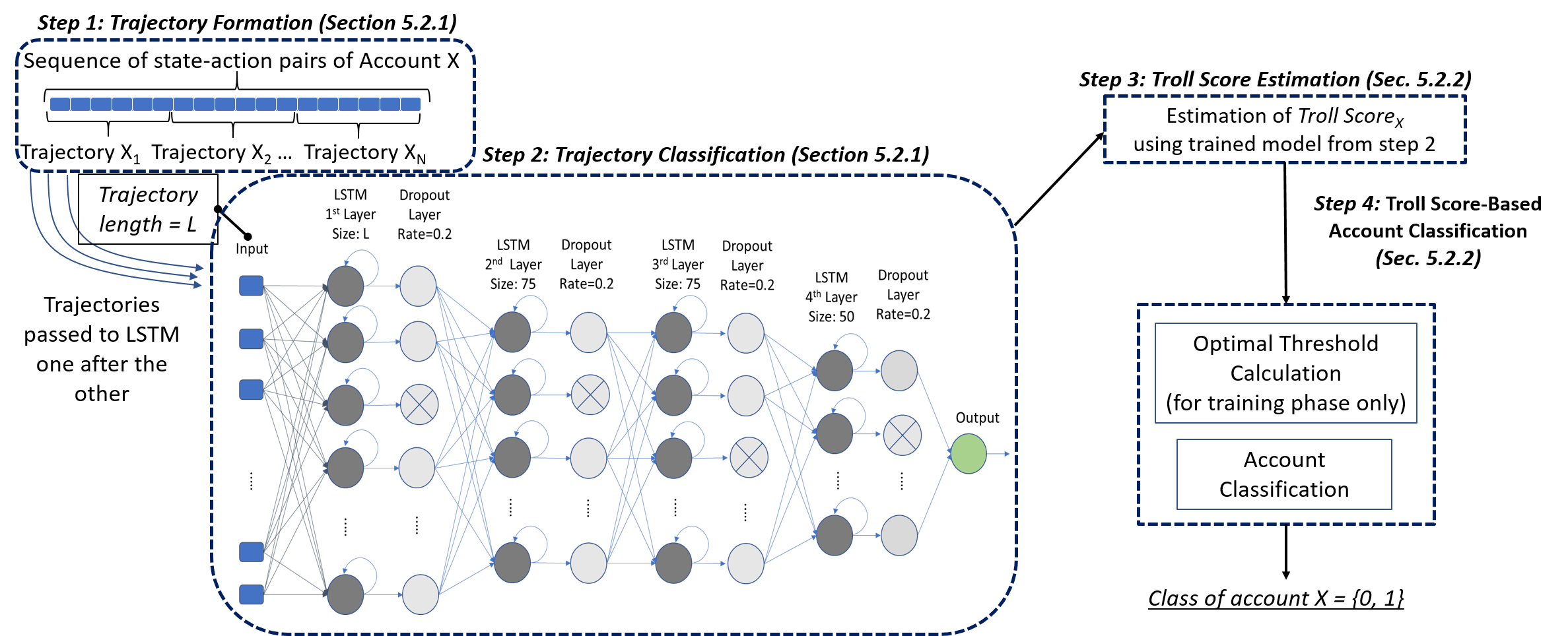}
    \caption{Architecture of the proposed approach for state-sponsored troll detection}
    \label{fig:overview}
\end{figure*}

\section{Methodology}
\label{method}

This Section details our proposed approach to identifying troll accounts. Section \ref{framework} provides an overview of our framework, while section \ref{troll_detection} describes the proposed methodology in more detail.

\subsection{Overall Framework}\label{framework}

The proposed framework for the identification of troll accounts (see Figure \ref{fig:overview}) consists of the following main steps:

\begin{itemize}
    \item \textbf{Step 1 --- Trajectory Formation}:     Creation of trajectories that represent portions of an account's sequence, where a trajectory is a time-sorted set of a pre-defined number of \emph{state-action pairs}. 
    \item \textbf{Step 2 --- LSTM-Based Trajectory Classification}: Classification of the trajectories into two classes, i.e., trolls' or users' trajectories. To perform this classification, we employ a Long Short-Term Memory (LSTM)-based classifier.
    \item \textbf{Step 3 --- Troll Score Estimation}: Computation of the \emph{Troll Score} as the ratio of the number of trajectories classified as belonging to a troll over the number of an account's trajectories. 
    \item \textbf{Step 4 --- Troll Score-Based Account Classification}: Classification of an account, i.e., \emph{troll} or \emph{user}, based on the computed \emph{Troll Score}.
\end{itemize}

Figure \ref{fig:overview} portrays an overview of our proposed framework. In  \emph{Step 1}, the sequence of online activities of an account is divided into several trajectories of state-action pairs of a pre-defined length. In \emph{Step 2}, an LSTM model fed with the extracted trajectories classifies every trajectory as either a troll trajectory or a user trajectory. Note that, in the training phase of the LSTM model, a trajectory extracted from a sequence of a troll is considered to have a label of a \emph{troll trajectory}, while that extracted from a sequence of a user is given the label of a \emph{user trajectory}. We employ deep learning and, more specifically, an LSTM model as we deal with time series data (i.e., sharing activities represented by state-action trajectories). 
After classifying the trajectories, a Troll Score of every account is computed as explained in \emph{Step 3}. Finally, in \emph{Step 4}, an account is classified as either a troll or a user based on the Troll Score. 

\subsection{Trolls Detection}
\label{troll_detection}
We now discuss in more detail our proposed approach for the identification of troll accounts. In Section \ref{LSTM}, we describe the LSTM-based model used to classify trajectories of state-action pairs, while in Section \ref{trollscoresection}, we introduce the \textit{Troll Score} metric and we detail the account classification approach. 
The code and scripts used to implement the methodological framework detailed below are freely available to the research community\footnote{https://github.com/FatimaEzzedinee/Exposing-Influence-Campaigns-in-the-Age-of-LLMs-A-Behavioral-Based-AI-Approach}.

\subsubsection{LSTM-Based Model for Trajectory Classification\\}\label{LSTM}

\textbf{Model Input.}
The LSTM model takes as input a trajectory of a given length, i.e., a pre-defined number $L$ of state-action pairs. To build these trajectories, two questions need to be answered:
    $i)$ How to choose the value of $L$?, and $ii)$ given a value of $L$, how to divide the sequence into trajectories?
    
\textbf{How to choose the value of $L$?} The choice of $L$ is not trivial, and it is, in fact, decisive. Considering a small value of $L$ is desired from a computational standpoint. However, a relatively small value for $L$ might not be enough for the model to distinguish troll trajectories from user trajectories. Conversely, considering a relatively large value of $L$ might not be feasible, as not every account might largely engage in online activities, i.e., not all the accounts might have a long sequence. 

\textbf{Given a value of $L$, how to divide the sequence into trajectories?} To form trajectories for trajectory classification, we consider non-overlapping parts of the sequence, i.e., we divide the sequence into $L$-long trajectories. For instance, for $L$ = 100 and a sequence composed of 200 state-action pairs, two trajectories, each of length $L$, are created and therefore considered for classification. Note that, for a given value of $L$, the number of trajectories to classify per account differs. A large $L$ leads to a low number of trajectories to classify, while a small $L$ leads to a high number of trajectories to classify. 
We will show the impact of this parameter in Section \ref{trajclassification} by performing sensitivity analysis and monitoring the performance of our model (in terms of several classification metrics) at varying $L$. 

\subsubsection{Troll Score-Based Classification of Accounts\\}\label{trollscoresection}

\textbf{Troll Score Definition.}
The rationale behind defining a \emph{Troll Score} is to have a measure to quantify the extent to which an account behaves like a troll in its online activity. We define the Troll Score of an account as the ratio between the number of trajectories classified as a troll trajectory by the LSTM model and the total number of trajectories of the account. The Troll Score, thus, ranges from zero to one, where a score close to one indicates a troll-like behavior, and a score close to zero denotes a user-like behavior. 
To compute the Troll Score of a given account, we consider a sliding window of length $L$ over the whole sequence of state-action pairs of every account under scrutiny. This approach allows each state-action pair to contribute to multiple trajectories.

\textbf{Troll Score for Account Classification.}
To classify accounts based on their Troll Score, a threshold to distinguish the two classes is required. 
We compute this threshold as follows. 
First, the Troll Score of a subset of the accounts is computed. Then, we iterate over all threshold values ranging from 0 to 1, with a step size of 0.02, reporting the performance of each threshold value in terms of AUC. That is, we consider each value to serve as a threshold and classify accounts accordingly, where each value will result in a different classification of the accounts and, hence, a distinct performance. Finally, we select the threshold that provides the best AUC. The same method can be applied to optimize other classification metrics, e.g., precision, recall, etc.
For our evaluations, we test our approach with a 10-fold cross-validation. 

\section{Implementation and Results}
In this Section, we first perform a sensitivity analysis comparing the performance of the LSTM-based trajectory classification approach in the case of trajectories composed of $(i)$ state-action pairs and $(ii)$ actions only. The objective here is to evaluate whether the states (i.e., received feedback) represent beneficial signals for the classification task. The rationale is to understand if the feedback received by trolls and users, along with the way they react to such stimuli, might allow us to better discern these two classes of accounts with respect to leveraging only their actions.
Then, we benchmark the trajectory classification approach based on LSTM, comparing its performance to those of off-the-shelf machine learning models. 

After validating the performance of the LSTM-based trajectory classification approach, we use it to compute the \textit{Troll Score} of the accounts under investigation. We then exploit the Troll Score to distinguish troll and user accounts, and we compare the classification performance to other approaches proposed in the literature. 
Moreover, based on the obtained results, we conduct an observational analysis to further investigate how the visited state-action pairs of trolls and users differ. 
Finally, we evaluate whether the proposed approach can generalize to the detection of entities operating in influence campaigns. We present the results of our approach on data from online discussions related to the U.S. 2020 Election on Twitter with the objective of distinguishing the drivers of influence campaigns from organic users.

\subsection{Trajectory Classification}\label{trajclassification}
This subsection presents the implementation details of our proposed LSTM model for trajectory classification and then discusses results comparing 
our proposed approach to off-the-shelf machine learning models. 

\subsubsection{Implementation Details}
Our proposed LSTM model for trajectory classification is composed of four LSTM layers, four dropout layers, and a dense layer. Each layer is followed by a Dropout Layer to reduce overfitting \cite{srivastava2014dropout}. The \emph{sigmoid} is used as an activation function for both the hidden and output layers.
We fine-tune the hyper-parameters of our model with a random search. We encode trajectories to be fed into the LSTM model using \emph{Label Encoding}, which consists of assigning an integer to each combination of state-action pairs. The entire model is trained by minimizing the binary cross-entropy with the Adam optimization algorithm \cite{nielsen2015neural}.

\subsubsection{State-Action vs. Action Sequences}
To perform trajectory classification, we rely on the historical data of the accounts. 
To evaluate our proposed approach, we build the trajectories of each of the accounts in two ways: $(i)$ considering state-action pairs sequences or $(ii)$ considering only the sequences of actions.
Further, we consider five different values of the trajectory length $L$, ranging from 50 to 200, as shown in Table \ref{data-description}. Note that the overall number of trajectories changes with the value of $L$. We report in Table \ref{data-description} the number of trajectories per every class of accounts in each of the cases (state-action pairs and only actions) and for all values of $L$. It is worth noting that the number of sequences based only on \emph{Actions} is much lower than those based on \emph{State-Action} pairs, as trajectories of \emph{Actions} are formed only by three sharing activities (\emph{tw}, \emph{rt}, and \emph{in}), and hence are much shorter than \emph{State-Action} trajectories. 
We train and test our LSTM model in both cases and report the results of a 10-fold stratified cross-validation. It should be noted that for the scenario with only \emph{Actions}, there is an imbalance between the number of sequences of trolls and users. To solve this issue, we employ the under-sampling technique \cite{liu2008exploratory, drumnond2003class}. 

\begin{table}[]
\centering
\begin{tabular}{l|l|l|l}
Input  & L & Trolls Trajectories & Users Trajectories  \\ \hline
\multirow{6}{*}{State-Action} & 50 & 64,984 & 102,101 \\ \cline{2-4} 
 & 65 & 49,975 & 78,424 \\ \cline{2-4} 
 & 80 & 40,587 & 63,643 \\ \cline{2-4}
 & 100 & 32,457 & 50,831 \\ \cline{2-4} 
 & 150 & 21,628 & 33,795 \\ \cline{2-4} 
 & 200 & 16,230 &  25,299 \\ \hline
\multirow{6}{*}{Action} & 50 & 7,709 & 93,721  \\ \cline{2-4} 
 & 65 & 5,927 & 72,091  \\ \cline{2-4} 
 & 80 & 4,802 & 58,593  \\ \cline{2-4}
 & 100 & 3,819 & 46,863  \\ \cline{2-4} 
 & 150 & 2,547 & 31,267 \\ \cline{2-4} 
 & 200 & 1,920 & 23,476 \\ 
\end{tabular}
\caption{Number of trajectories of trolls and users for each trajectory length $L$}
\label{data-description}
\end{table}

\begin{figure*}[t]

\subfigure[]{
    \label{stateactionsvsactions1}
    \begin{tikzpicture}
        \begin{axis}[
        xlabel={\small{Trajectory Length}},
        xtick = {50, 65, 80, 100, 150, 200},
        grid=both,
        grid style={line width=.1pt, draw=gray!10},
        major grid style={line width=.2pt,draw=gray!50},
        xmin = 45, 
        ymax = 100,
        xmax = 200,
        ylabel={\%}, 
        width=0.45\textwidth,
        height=0.4\textwidth,
	    legend style={draw=none, at={(0.5,1.33)},anchor=north,legend columns=2}
        ]
        \addplot[draw = black, mark = x, thick] plot coordinates {(200,99.68) (150,99.7) (100, 99.57) (80, 99.5) (65, 99.4) (50,99.3)};
        \addlegendentry{\small{AUC State-Action}}
        
        \addplot[dashed, draw = black, mark = x, thick] plot coordinates {(200,98.5) (150,98.3) (100, 98.1) (80,98) (65, 97) (50, 97)};
        \addlegendentry{\small{Accuracy State-Action}}
        
        \addplot[draw = blue, mark = x, thick] plot coordinates {(200,81.4) (150,81.25)  (100, 81.9) (80, 81.6) (65, 81.7) (50,81.5)};
        \addlegendentry{\small{AUC Action}}
        
        \addplot[dashed, draw = blue, mark = x, thick] plot coordinates {(200,92.5) (150, 92.6)  (100, 92.4) (80, 92.5) (65, 92.4) (50, 92.1)
        };
        \addlegendentry{\small{Accuracy Action}}
    \end{axis}
    \end{tikzpicture}
   }
\subfigure[]
{
    \label{stateactionsvsactions2}
    \begin{tikzpicture}
        \begin{axis}[
        xlabel={\small{Trajectory Length}},
        xtick = {50, 65, 80, 100, 200},
        grid=both,
        grid style={line width=.1pt, draw=gray!10},
        major grid style={line width=.2pt,draw=gray!50},
        xmin = 45, 
        ymax = 101,
        ylabel={\%}, 
        width=0.45\textwidth,
        height=0.4\textwidth,
	    legend style={draw=none, at={(0.5,1.45)},anchor=north,legend columns=2}
        ]
        \addplot[draw = black, mark=o, thick, mark options={fill=white,scale=1}] plot coordinates {(200,98.5) (150, 98.31) (100, 98.1) (80, 97.5) (65, 97) (50, 96)};
        \addlegendentry{\small{Precision State-Action}}
        
        \addplot[draw = blue, mark=o, thick, mark options={fill=white,scale=1}] plot coordinates {(200,59.8) (150,33.33)   (100, 16.5) (80,11) (65,11) (50,10)};
        \addlegendentry{\small{Precision Action}}
        
        \addplot[dashed, draw = black, mark=o, thick, mark options={fill=white,scale=1}] plot coordinates {(200,97.7) (150, 97.56) (100, 97.1) (80, 97.1) (65, 97) (50, 97)};
        \addlegendentry{\small{Recall State-Action}}

        \addplot[dashed, draw = blue, mark=o, thick, mark options={fill=white,scale=1}] plot coordinates {(200,1.8) (150, 1.3)   (100, 0.2) (80,0) (65,0) (50,0) };
        \addlegendentry{\small{Recall Action}}
        
        \addplot[dotted,draw = black, mark=o, thick, mark options={fill=white,scale=1}] plot coordinates {(200,98.1) (150,97.9) (100, 97.1) } (80, 97.2) (65, 97) (50, 96.4);
        \addlegendentry{\small{F1 State-Action}}
        
        \addplot[dotted, draw = blue, mark=o,thick, mark options={fill=white,scale=1}] plot coordinates {(200,3.5) (150, 2) (100, 0.5) (80,0) (65,0) (50,0)};
        \addlegendentry{\small{F1 Action}}
    \end{axis}
    \end{tikzpicture}
    }
\caption{ (a) {\small{AUC and Accuracy of the LSTM-based sequence classification approach with trajectories composed of state-action pairs or actions only}}, and (b) \small{Precision, Recall, and F1-Score of the LSTM-based sequence classification approach with trajectories composed of state-action pairs or actions only}}
\end{figure*}
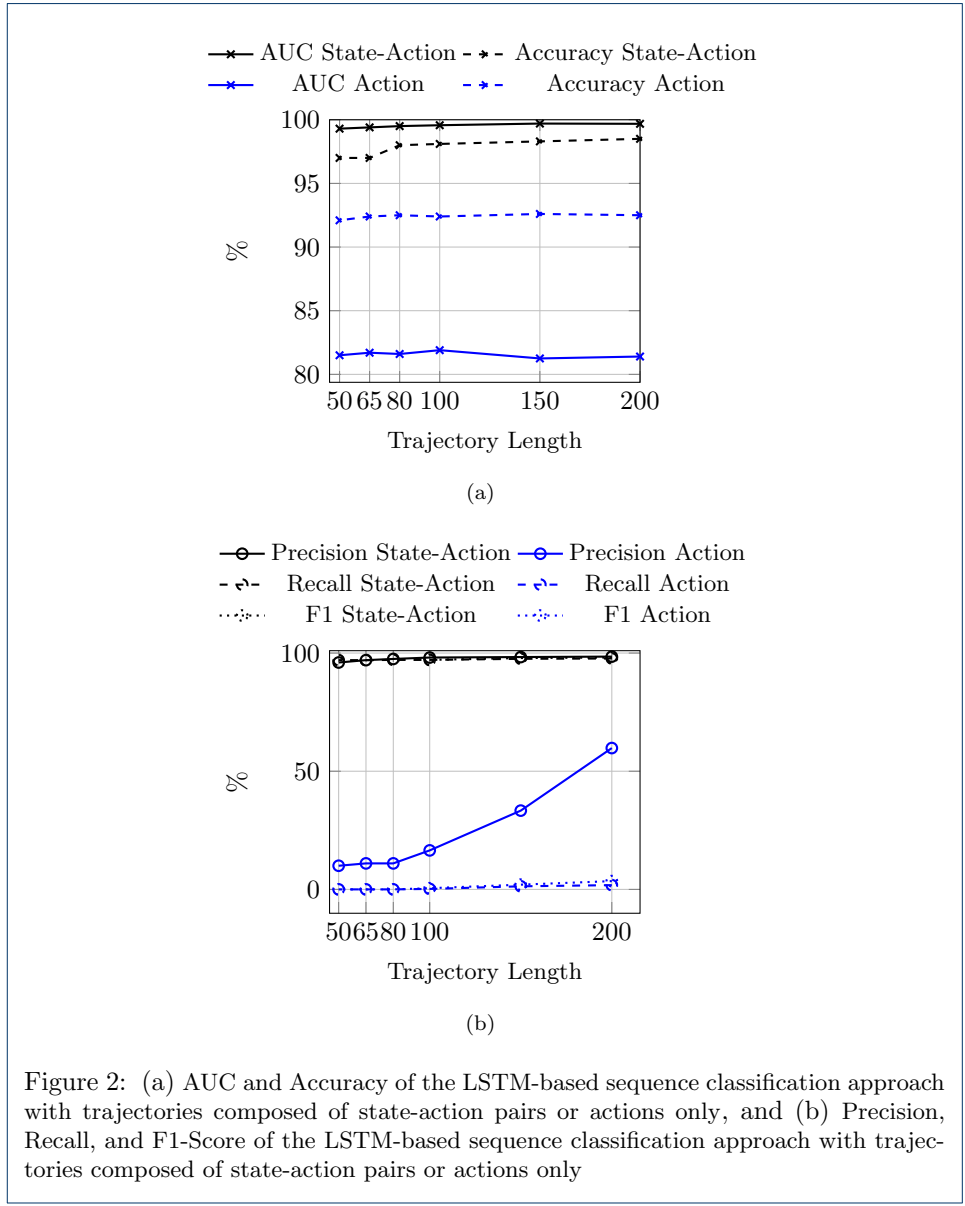

Figure \ref{stateactionsvsactions1} shows the AUC and accuracy of the LSTM-classifier with trajectories composed of $(i)$ \emph{State-Action} pairs and $(ii)$ \emph{Actions} only as functions of the trajectory length $L$. Results show that both classification metrics are higher when \emph{State-Action} pairs are considered with respect to sequences of \emph{Actions} for every value of $L$. Specifically, when sequences are composed of \emph{State-Action} pairs, the AUC is about 99\%, significantly higher than when considering \emph{Actions} only (AUC around 82\%). This is also reflected in the classification performance in terms of accuracy (97\% with \emph{State-Action} pairs vs. 92\% with \emph{Actions} only). This suggests that the \emph{states} (i.e., feedback from other users) represent beneficial signals for the accounts classification task as, along with the \emph{actions}, allow achieving a better distinction between trolls and users if compared to the results achieved when considering trajectories composed of actions only. From Fig. \ref{stateactionsvsactions1}, we also note that varying the value of $L$ has a minimal impact on the model's performance in both cases. In fact, the AUC of the LSTM-based approach reaches 99\% even when $L$ is relatively short (e.g., $L$ = 50). This finding shows that the proposed model (when considering \emph{State-Action}) has the ability to correctly identify trolls' and users' sequences even when little information about accounts’ online activity is available. This surprisingly high classification accuracy is probably due to the nature of our experimental design, which focuses on distinguishing the activity sequences of organic users from those of troll accounts. The latter, performing their agenda regardless of others' feedback \cite{luceri2020detecting}, present activity patterns naturally different with respect to legitimate users. While these differences might appear explicit in such a high-quality dataset, we do not expect the same results in a more challenging scenario (see Section \ref{sectionsuspendedusers}).


We further evaluate the performance of our model by considering other classification metrics such as Precision, Recall, and F1 score. Figure \ref{stateactionsvsactions2} depicts these metrics for the different values of $L$. Results show that the model has a nearly perfect performance with \emph{State-Action} considering all metrics (around 98\%) with a slight increase in performance as $L$ increases. On the contrary, the LSTM classifier with \emph{Actions} sequences suffers from low performance, nearly 0\%, in terms of Recall and F1, while precision ranges between 10\% and 60\% and increases as $L$ increases. 
This further confirms our previous intuition in complementing actions with the feedback the accounts receive from the environment (i.e., states). Indeed, these results show that the states are essential to accurately classify the two classes of accounts, which suggests that trolls might react to online feedback differently from non-trolls. 
In the rest of our analysis, we will consider the model based on \emph{State-Action} pairs, given that it has been shown to be effective in discriminating between users and trolls.

\begin{figure*}[t]
\hspace{-20pt}
\subfigure[]
{
    \centering
    \begin{tikzpicture}
        \begin{axis}[
        xlabel={Sequence length L},
        xtick={100, 200},
        grid=both,
        grid style={line width=.1pt, draw=gray!10},
        major grid style={line width=.2pt,draw=gray!50},
        xmin = 90, 
        ymin = 90,
        ymax = 100,
        ylabel near ticks,
        ylabel={\small{AUC}}, 
        width=0.34\textwidth, 
        height=.39\textwidth,
	    legend style={draw=none, at={(0.5,1.2)},anchor=north,legend columns=4}
        ]
        \addplot[draw = black, mark=o, ultra thick] plot coordinates {(100,99.5) (200,99.6)};
        
        \addplot[draw = blue, mark=o, ultra thick] plot coordinates {(100,97) (200,97)};
       
        \addplot[draw = gray, mark=o, ultra thick] plot coordinates {(100,96) (200,96)};
        
        \addplot[dotted, draw = black, mark=o, ultra thick] plot coordinates {(100,95) (200,96)};
        
        \addplot[dotted, draw = blue, mark=o, ultra thick] plot coordinates {(100,98) (200,98)};

        \addplot[dotted, draw = gray, mark=o, ultra thick] plot coordinates {(100,93) (200,92)};
        
        \addplot[draw = brown, mark=o, ultra thick] plot coordinates {(100,98) (200,97)};
        \legend{{\fontsize{7}{7}\selectfont{LSTM}}, {\fontsize{7}{7}\selectfont{Ada Boost}}};
 
    \end{axis}
    \end{tikzpicture}
}
\hspace{-20pt}
\subfigure[]
{
    \centering
    \begin{tikzpicture}
        \begin{axis}[
        xlabel={Sequence length L},
        xtick={100, 200},
        grid=both,
        grid style={line width=.1pt, draw=gray!10},
        major grid style={line width=.2pt,draw=gray!50},
        xmin = 90, 
        ymin = 90,
        ymax = 100,
        ylabel near ticks,
        ylabel={\small{Recall}}, 
        width=0.34\textwidth, 
        height=0.39\textwidth,
	    legend style={draw=none, at={(0.47,1.2)},anchor=north,legend columns=4}
        ]
        \addplot[draw = black, mark=o, ultra thick,forget plot] plot coordinates {(100,97.7) (200,97.1)};
        
        \addplot[draw = blue, mark=o, ultra thick, forget plot] plot coordinates {(100,94) (200,94)};
        
        \addplot[draw = gray, mark=o, ultra thick] plot coordinates {(100,94) (200,92)};
        
        \addplot[dotted, draw = black, mark=o, ultra thick] plot coordinates {(100,90) (200,91)};
        
        \addplot[dotted, draw = blue, mark=o, ultra thick] plot coordinates {(100,93.1) (200,92.3)};

        \addplot[dotted, draw = gray, mark=o, ultra thick] plot coordinates {(100,94) (200,93)};
        
        \addplot[draw = brown, mark=o, ultra thick] plot coordinates {(100,92) (200,92)};
        \legend{{\fontsize{7}{7}\selectfont{ANN}}, {\fontsize{7}{7}\selectfont{KNN}}};
    \end{axis}
    \end{tikzpicture}
}
\hspace{-20pt}
\subfigure[]
{
    \centering
    \begin{tikzpicture}
        \begin{axis}[
        xlabel={Sequence length L},
        xtick={100, 200},
        grid=both,
        grid style={line width=.1pt, draw=gray!10},
        major grid style={line width=.2pt,draw=gray!50},
        xmin = 90, 
        ymin = 90,
        ymax = 100,
        ylabel near ticks,
        ylabel={\small{Precision}}, 
        width=0.34\textwidth,  
        height=.39\textwidth,
	    legend style={draw=none, at={(0.45,1.2)},anchor=north,legend columns=4}
        ]
        \addplot[draw = black, mark=o, ultra thick, forget plot] plot coordinates {(100,98.1) (200,98.5)};
        
        \addplot[draw = blue, mark=o, ultra thick, forget plot] plot coordinates {(100,90) (200,90)};

        \addplot[draw = gray, mark=o, ultra thick, forget plot] plot coordinates {(100,89) (200,89)};
        
        \addplot[dotted, draw = black, mark=o, ultra thick, forget plot] plot coordinates {(100,91) (200,91)};
        
        \addplot[dotted, draw = blue, mark=o, ultra thick] plot coordinates {(100,92) (200,91)};

        \addplot[dotted, draw = gray, mark=o, ultra thick] plot coordinates {(100,91) (200,89)};
        
        \addplot[draw = brown, mark=o, ultra thick] plot coordinates {(100,89) (200,89)};
        \legend{{\fontsize{7}{7}\selectfont{SVM}}, {\fontsize{7}{7}\selectfont{Decision trees}}};      
    \end{axis}
    \end{tikzpicture}
}
\hspace{-20pt}
\subfigure[]
{
    \centering
    \begin{tikzpicture}
        \begin{axis}[
        xlabel={Sequence length L},
        xtick={100, 200},
        grid=both,
        grid style={line width=.1pt, draw=gray!10},
        major grid style={line width=.2pt,draw=gray!50},
        xmin = 90, 
        ymin = 90,
        ymax = 100,
        ylabel near ticks,
        ylabel={\small{F1-Score}}, 
        width=0.34\textwidth, 
        height=0.39\textwidth,
	    legend style={draw=none, at={(0.5,1.2)},anchor=north,legend columns=4}
        ]
        \addplot[draw = black, mark=o, ultra thick, forget plot] plot coordinates {(100,97.6) (200,98.1)};
        
        \addplot[draw = blue, mark=o, ultra thick, forget plot] plot coordinates {(100,91) (200,91)};
        
        \addplot[draw = gray, mark=o, ultra thick, forget plot] plot coordinates {(100,91) (200,90)};
        
        \addplot[dotted, draw = black, mark=o, ultra thick, forget plot] plot coordinates {(100,91) (200,91)};
        
        \addplot[dotted, draw = blue, mark=o, ultra thick, forget plot] plot coordinates {(100,92) (200,91)};

        \addplot[dotted, draw = gray, mark=o, ultra thick, forget plot] plot coordinates {(100,92) (200,91)};
        
        \addplot[draw = brown, mark=o, ultra thick] plot coordinates {(100,92) (200,92)};
        \legend{{\fontsize{7}{7}\selectfont{Logistic Regression}}}; 
    \end{axis}
    \end{tikzpicture}
    }
    \caption{Comparison between the proposed LSTM-based approach and off-the-shelf machine learning models for trajectory classification in terms of (a) AUC, (b) Recall, (c) Precision, and (d) F1-score}
\label{lstmvsofftheshelf}
\end{figure*}

\subsubsection{LSTM vs. Off-the-Shelf Machine Learning Models}
\label{lstm_vs_offtheshelf}
We now compare the performance of our proposed LSTM-based model for trajectory classification to those of off-the-shelf machine learning models. Specifically, we consider Support Vector Machine (SVM), Artificial Neural Network (ANN),  Decision Tree, Ada Boost, Logistic Regression, and K-Nearest Neighbors (KNN). Fig. \ref{lstmvsofftheshelf} shows AUC, Recall, Precision, and F1-score of these models for $L$ = 100 and $L$ = 200. For all metrics, and for both values of $L$, all machine learning models show a promising performance (e.g., AUC of 92\% and higher), whereas our proposed LSTM-based approach achieves a near-optimal performance, thus outperforming all other models. More specifically, the LSTM-based approach shows 98\% for Precision and Recall, significantly higher than that of other approaches, which do not exceed 92\%. This indicates that trolls' and users' trajectories embed distinguishable patterns, which are more easily identifiable by models conceived to work with sequential data, such as LSTM, than other general-purpose machine learning models.  

\subsection{Troll Score for Account Classification}
We now focus our discussion on the classification of accounts. In this scenario, we build trajectories by sliding a window of length $L$ with a step of one element over the entire sequence of state-action pairs. Note that in this case, two consecutive trajectories overlap by $L-1$ elements. Indeed, the same state-action pair is considered into several trajectories, depending on its position within the sequence. 
As previously discussed, the classification of the accounts is based on the \textit{Troll Score} metric. The latter is computed for every account, leveraging the classification of its trajectories extracted with a sliding window. Finally, the Troll Score of the account under scrutiny is compared to a \emph{Troll Score threshold} (see Section \ref{method}.2.2) to assign the account to one of the two classes of accounts (troll vs. user). 

\subsubsection{Troll Score of User and Troll Accounts\\}

\begin{figure}[t] 
	\centering	
	{
		\begin{tikzpicture}
		\begin{axis}[ 
		ylabel near ticks,
        width=0.45\textwidth, 
        height=0.4\textwidth,
		ylabel={Density},
		xlabel={Troll Score},
		xticklabel style={
			/pgf/number format/fixed,
			/pgf/number format/precision=2		
		},
		xmin=0,
		xmax=1.0,
		xtick={0,0.2, 0.4, 0.6, 0.8, 1},
		ymin=0, 
		ymax=1.0,
		ytick={0,0.2, 0.4, 0.6, 0.8, 1.0},
		grid=both,
        grid style={line width=.1pt, draw=gray!10},
        major grid style={line width=.2pt,draw=gray!50},
		]
		\pgfplotstableread{Results/sample_x_y_users_200.txt}{\comp};
		\addplot [color=blue, ultra thick, mark size=1.3, ] table [x index=1, y index=2] {\comp}; 
		
		\pgfplotstableread{Results/sample_x_y_trolls_200.txt}{\comp};
		\addplot [color=black, ultra thick, mark size=1.3] table [x index=1, y index=2]{\comp};
		\legend{Users, Trolls} 
		\end{axis}
		\end{tikzpicture}
	}
	\caption{Troll Score CDF of user and troll accounts with $L$ = 200}	
	\label{trollscore200}
\end{figure}

We first analyze the Troll Score computed for each account in our dataset. Figure \ref{trollscore200} shows the Cumulative Distribution Function (CDF) of the Troll Score of user and troll accounts for $L$ = 200\footnote{We omit showing the case with $L$ = 100 as results present same insights as for $L$ = 200.}. 
Numerical results show that the CDF of the Troll Score of users shows a logarithmic growth (the CDF reaches 0.95 for a Troll Score of 0.02). This indicates that most of the users (95\% of them) have an almost-zero Troll Score while the rest (remaining 5\%) have a Troll Score of at most 0.2, which can be considered considerably low. On the contrary, the CDF related to the trolls follows exponential growth: It increases very slowly for low values of Troll Score and then increases exponentially for higher values of Troll Score. In particular, results show that 80\% of the trolls have a Troll Score of at least 0.8. On the one hand, this suggests that a significant portion of the trolls are characterized by a high Troll Score. On the other hand, this indicates that a limited yet considerable portion of trolls have a relatively low Troll Score, e.g., 10\% have a Troll Score of at most 0.5. This finding demonstrates that some of the trolls under analysis, specifically those characterized by a low Troll Score, exhibit user-like behavior. To inspect this in more detail, we perform an analysis to observe the visited state-action pairs of all accounts. We present this observational analysis in the following section.

\subsubsection{Behavioral Clustering and Troll Score}
This section presents an analysis conducted to analyze the behavior of troll and user accounts in terms of their visited state-action pairs, i.e., state and action pairs that are present in the account's trajectory.
The rationale behind this analysis is to examine whether trolls and users can be distinguished by looking at their visited state-action pairs. Specifically, we evaluate whether the different classes of accounts are grouped into distinct clusters and, if clusters are observed, their relation with the Troll Score.

\begin{table}[b]
\centering
\begin{tabular}{l|l|l|l}
        & Cluster 1 & Cluster 2 & Cluster 3 \\ \hline
 Users  & 977 (49.5\%) & 603 (30.43\%) & 401 (20.24\%)  \\ \hline
 Trolls & 256 (75.07\%) & 65 (19.06\%) & 20 (5.86\%)
\end{tabular}
\caption{Distribution of user and troll accounts in the three naturally-emerging clusters}
\label{tab:Clusters}
\end{table}

\begin{figure*}[t]
\centering
\begin{subfigure}[]
    \centering
    \includegraphics[width=150pt, height=150pt]{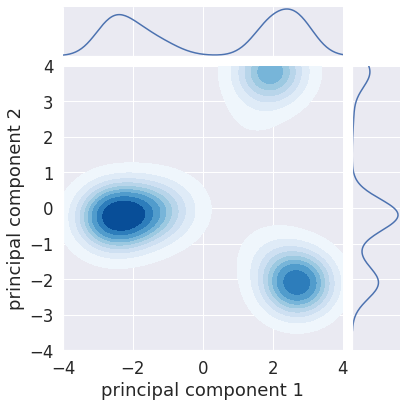}
\end{subfigure}
\begin{subfigure}[]
    \centering
    \includegraphics[width=150pt, height=150pt]{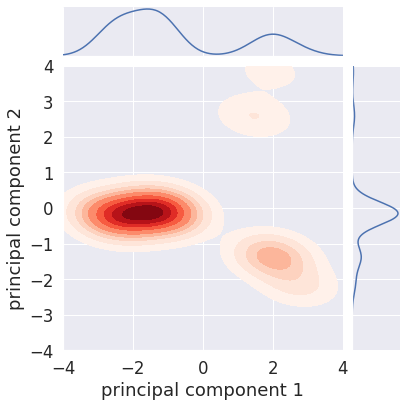}
\end{subfigure}
\begin{subfigure}[]
    \centering
    \includegraphics[width=150pt, height=150pt]{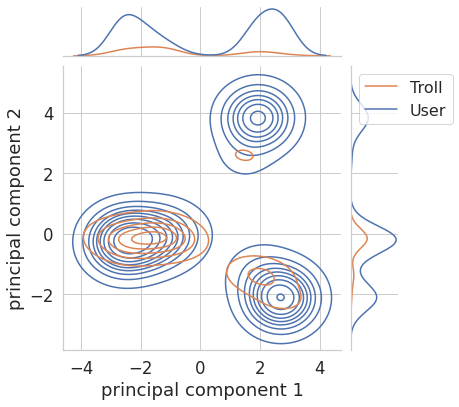}
\end{subfigure}
\caption{Clusters of accounts based on the visited state-action pairs for (a) users, (b) trolls, and (c) users and trolls combined}
\label{clustering}
\end{figure*}

\begin{figure*}[]
\hspace{-0.35cm}
\subfigure[Troll Score CDF for Cluster 1]
{
		\begin{tikzpicture}
		\begin{axis}[ 
		ylabel near ticks,
        width=0.35\textwidth, 
        height=0.35\textwidth,
		ylabel={Density},
		xlabel={Troll Score},
		xticklabel style={
			/pgf/number format/fixed,
			/pgf/number format/precision=2		
		},
		xmin=0,
		xmax=1.0,
		xtick={0,0.2, 0.4, 0.6, 0.8, 1},
		ymin=0, 
		ymax=1.0,
		ytick={0,0.2, 0.4, 0.6, 0.8, 1.0},
		grid=both,
        grid style={line width=.1pt, draw=gray!10},
        major grid style={line width=.2pt,draw=gray!50},
		]
		\pgfplotstableread{Results/sample_x_y_users_200_C1.txt}{\comp};
		\addplot [color=blue, ultra thick, mark size=1.3, ] table [x index=1, y index=2] {\comp}; 
		
		\pgfplotstableread{Results/sample_x_y_trolls_200_C1.txt}{\comp};
		\addplot [color=black, ultra thick, mark size=1.3] table [x index=1, y index=2]{\comp};
		\legend{Users, Trolls} 
		\end{axis}
		\end{tikzpicture}
	}
\hspace{-0.35cm}
\subfigure[Troll Score CDF for Cluster 2]
{
		\begin{tikzpicture}
		\begin{axis}[ 
		ylabel near ticks,
        width=0.35\textwidth, 
        height=0.35\textwidth,
		ylabel={Density},
		xlabel={Troll Score},
		xticklabel style={
			/pgf/number format/fixed,
			/pgf/number format/precision=2		
		},
		xmin=0,
		xmax=1.0,
		xtick={0,0.2, 0.4, 0.6, 0.8, 1},
		ymin=0, 
		ymax=1.0,
		ytick={0,0.2, 0.4, 0.6, 0.8, 1.0},
		grid=both,
        grid style={line width=.1pt, draw=gray!10},
        major grid style={line width=.2pt,draw=gray!50},
		]
		\pgfplotstableread{Results/sample_x_y_users_200_C2.txt}{\comp};
		\addplot [color=blue, ultra thick, mark size=1.3, ] table [x index=1, y index=2] {\comp}; 
		
		\pgfplotstableread{Results/sample_x_y_trolls_200_C2.txt}{\comp};
		\addplot [color=black, ultra thick, mark size=1.3] table [x index=1, y index=2]{\comp};
		\end{axis}
		\end{tikzpicture}
	}
	\hspace{-0.35cm}
\subfigure[Troll Score CDF for Cluster 3]
{
		\begin{tikzpicture}
		\begin{axis}[ 
		ylabel near ticks,
        width=0.35\textwidth, 
        height=0.35\textwidth,
		ylabel={Density},
		xlabel={Troll Score},
		xticklabel style={
			/pgf/number format/fixed,
			/pgf/number format/precision=2		
		},
		xmin=0,
		xmax=1.0,
		xtick={0,0.2, 0.4, 0.6, 0.8, 1},
		ymin=0, 
		ymax=1.0,
		ytick={0,0.2, 0.4, 0.6, 0.8, 1.0},
		grid=both,
        grid style={line width=.1pt, draw=gray!10},
        major grid style={line width=.2pt,draw=gray!50},
		]
		\pgfplotstableread{Results/sample_x_y_users_200_C3.txt}{\comp};
		\addplot [color=blue, ultra thick, mark size=1.3, ] table [x index=1, y index=2] {\comp}; 
		
		\pgfplotstableread{Results/sample_x_y_trolls_200_C3.txt}{\comp};
		\addplot [color=black, ultra thick, mark size=1.3] table [x index=1, y index=2]{\comp};
		\end{axis}
		\end{tikzpicture}
	}
	\caption{Troll Score of accounts in Cluster 1 (a), 2 (b), and 3 (c) with $L$ = 200}
	\label{trollscoreclusters}
\end{figure*}

\paragraph{Clustering based on visited state-action pairs.} 
We perform user clustering considering the visited state-action pairs as a set of features, which, therefore, consists of the 11 possible combinations of states and actions. For every account, the values of the features are either set to 1, if the account visited a state-action pair, or to 0 otherwise. We use the Principal Component Analysis (PCA) to perform a dimensionality reduction, and we observe whether distinct clusters naturally emerge.
In Figure \ref{clustering}, we show the results of a PCA with two components, and we observe three clusters populated by both user and troll accounts.
Figure \ref{clustering}(a) shows that users divide into three distinct clusters with slightly different distributions. Similarly, Fig. \ref{clustering}(b) shows three clusters for trolls, in which one cluster embeds the majority of troll accounts. For a better comparison, we display in figure \ref{clustering}(c) the joint plot of users and trolls combined, from which we can appreciate the three distinct \emph{behavioral clusters}, where trolls and users coexist.

Table \ref{tab:Clusters} reports the number of trolls and users present in each of the three clusters. Cluster 1 refers to the largest cluster (bottom left in Figure \ref{clustering}), Cluster 2 refers to the cluster on the bottom right, and Cluster 3 refers to the one on the top right. By observing the visited state-action pairs, the three clusters can be differentiated as follows:

\begin{itemize}
    \item Cluster 1: Accounts in this cluster tweet or stay silent with any received feedback; 
    \item Cluster 2: Accounts in this cluster tweet when no feedback is provided by the environment (state NT) or they retweet with any received feedback; 
    \item Cluster 3: Accounts in this cluster tweet or interact with others (replying to or mentioning other accounts) with any received feedback. 
\end{itemize}

Our findings are consistent with \cite{luceri2020detecting} and show that most trolls (trolls in Clusters 1 and 2) tweet and retweet regardless of the received feedback, while a very small percentage of them (Cluster 3) participate in discussions by means of replies and mentions. This behavior is significantly different with respect to that of user accounts, as they are more evenly distributed among the three clusters and tend to join and participate more in discussions.

\paragraph{Troll Score per Cluster.} To further inspect the behavior of users and trolls in the three clusters, we evaluate the Troll Score of the accounts belonging to each of these clusters separately. Figures \ref{trollscoreclusters}(a)-(c) show the CDF of the Troll Score of accounts in Clusters 1-3, respectively. Figures show that the CDF of the Troll Score of users in all clusters has an \emph{exponential} shape, indicating that most users have a near-zero Troll Score. In contrast, the CDF of the Troll Score of trolls in every cluster shows a \emph{logarithmic} shape, suggesting that most of the trolls have a high Troll Score. However, the CDF of trolls within every cluster has notable differences.

For instance, in Clusters 1 and 2, most trolls are characterized by high Troll Scores (e.g., only about 20\% of trolls have a Troll Score below 0.8), while in Cluster 3, 60\% of trolls have a Troll Score below 0.8. This suggests that trolls belonging to Cluster 3 have a relatively low Troll Score, indicating that a relevant part of their trajectories is classified as user trajectories. This is likely due to the peculiar behavior of the small set of troll accounts (5\%) belonging to Cluster 3.
Based on this observation, we argue that our LSTM model identifies specific patterns of activities of trolls in Clusters 1 and 2, as they represent the most populated clusters (95\% of the accounts). 

\subsubsection{Troll Score-Based Classification vs. Existing Approaches\\} 
In this Section, we evaluate the classification performance of our method against other existing approaches.
In our proposed methodology, the training set is used to find the \textit{optimal} Troll Score threshold, while the test set is used to evaluate the classification. The value of $L$ is set to 200, based on the results in Section \ref{lstm_vs_offtheshelf}. We evaluate the performance of our proposed Troll Score-based approach by comparing it to $(i)$ a behavioral approach based on Inverse Reinforcement Learning (IRL) \cite{luceri2020detecting}, which is the only other language-agnostic solution that relies solely on the sequences of accounts' sharing activity to detect trolls.; and $(ii)$ two linguistic approaches \cite{addawood2019linguistic,im2020still} that use the text of the shared content to extract linguistic features for identifying troll accounts.


\begin{table}[t]
\centering

\begin{tabular}{l|c|c|c|c}
\hline
  & \multicolumn{2}{c|}{Behavioral Approaches} & \multicolumn{2}{c}{Linguistic Approaches} \\
Metric    & Troll Score-based & IRL-based \cite{luceri2020detecting} & Addawood et al. \cite{addawood2019linguistic} & Im et al. \cite{im2020still} \\ \hline
Accuracy  & \textbf{90.6}     &  83.0     & 98.9      & 97.7\\ \hline
AUC       & \textbf{90.5}     & 89.1      & 99.8       & 99.7 \\ \hline
Precision & \textbf{90.1}     & 84.0      & 98.5       & 97.5 \\ \hline
Recall    & \textbf{89.6}     &  85.0     & 97.3       & 97.5 \\ \hline
F1-Score  & \textbf{89.7}     & 84.5      & 98.9       & 97.5  \\ 
\end{tabular}
\caption{Comparison between behavioral and linguistic approaches for troll account identification}
\label{comparison}
\end{table}

Table \ref{comparison} displays the AUC, recall, precision, and F1-score of the behavioral and linguistic approaches tested through a stratified 10-fold cross-validation. 
Two facts are worth noting. First, the Troll Score-based solution outperforms the other behavioral approach \cite{luceri2020detecting} with an AUC of 90.6\%, a precision of 90.1\%, and a recall of 89.6\%.
Second, the linguistic approaches achieve high classification accuracy, with all the metrics close to 100\%. As suggested by prior research \cite{ionin2008sources,nicolai2014does}, the high classification accuracy of linguistic approaches is likely due to patterns of English misuse of content and function words among non-native English speakers, such as Russian trolls, which allows for identifying their original spoken language even when the speaker is fluent \cite{im2020still}. However, it is noteworthy that our Troll Score-based approach, which does not have access to the textual content of shared messages, approaches the performance of linguistic approaches that leverage message content. This represents a valuable advantage of behavioral models, especially considering the increasing capabilities of LLMs to refine and correct human-generated text. In the future, it will be crucial to develop models that use behavioral cues to detect deceptive activities, as distinguishing between content generated or revised by LLMs and human actors is becoming increasingly challenging \cite{mitrovic2023chatgpt,menczer2023addressing}. Based on these premises, in the next Section, we propose to investigate the generalizability of our approach to identifying drivers of information operations,\footnote{https://help.twitter.com/en/rules-and-policies/platform-manipulation} which might employ a more sophisticated combination of human curation, automation through social bots, and LLM-powered content generation \cite{nwala2022general,yang2023anatomy}.

\subsection{Identification of Drivers of Information Operations}\label{sectionsuspendedusers}

\begin{figure}
    \centering
    \includegraphics[width=350pt, height=170pt]{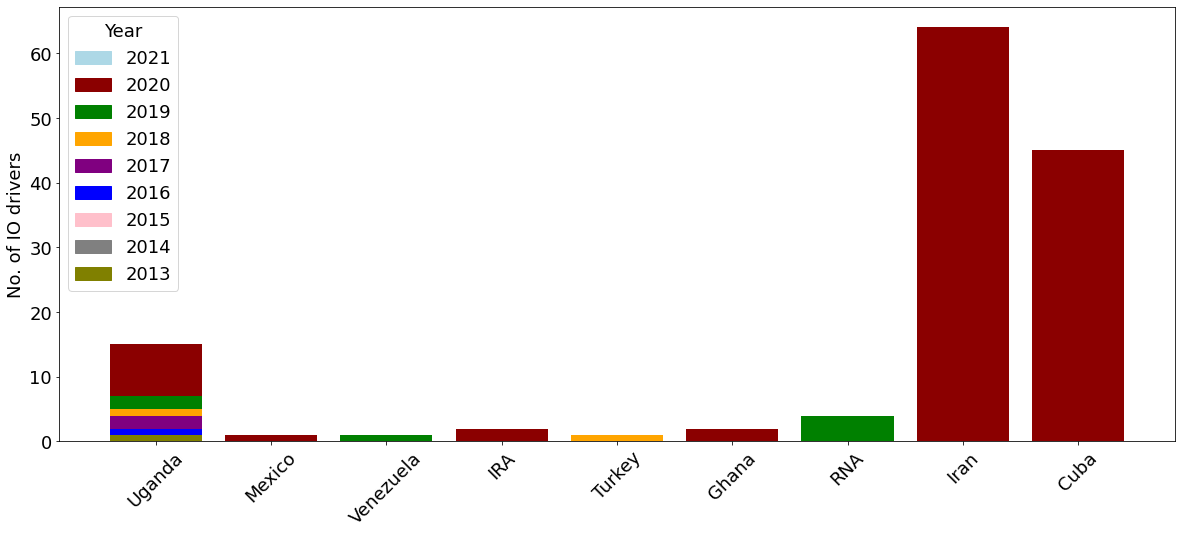}
    \caption{IO drivers during the 2020 U.S. Presidential Election}
    \label{fig:my_label}
\end{figure}

In this Section, we investigate whether the proposed approach can effectively identify a diverse set of actors involved in influence campaigns. Similarly to Nwala et al. \cite{nwala2022general}, we refer to these malicious entities as ``drivers of information operations" (\emph{IO drivers} for short). 
Specifically, we evaluate our approach's effectiveness in detecting IO drivers involved in discussions related to the 2020 U.S. Presidential election on Twitter. To conduct our analysis, we use the dataset made available by Chen et al. \cite{chen2021election2020}, which includes tweets that mention specific election-related keywords. We identify IO drivers in this dataset by consulting the list of accounts involved in state-linked information operations released by Twitter \cite{gadde2020additional}.

 The rationale behind interleaving these two data sources is twofold. First, we are interested in observing orchestrated interference campaigns originating from different countries, similar to Russian meddling in the 2016 U.S. election. As shown in Figure \ref{fig:my_label}, we uncovered influence efforts from a large set of information operations, with the majority of IO drivers operating in campaigns linked to Iran, Cuba, and Uganda. Notably, although these accounts originate from different countries, they almost exclusively share content in English, with 90\% of their tweets being in English.
Second, Twitter IO archive \cite{gadde2020additional} includes only messages shared by IO drivers during their lifespan on Twitter, which requires collecting another dataset to produce the negative class. This, in turn, may introduce biases in the data collection, primarily due to the selection of organic users and their potentially diverse activity on Twitter. Additionally, Twitter data only incorporates IO drivers' \emph{active online activities}, whereas our methodology leverages the feedback received by these accounts (\emph{passive online activities}). However, as these accounts have been suspended from Twitter, it is now impossible to collect organic users' interactions with IO drivers. Overall, the proposed study case represents a more challenging scenario with respect to the detection of IO drivers using Twitter data, as we do not benefit from the whole history of IO drivers' sharing activity on Twitter, but we can only observe their actions and received interactions during the 2020 U.S. election.

\begin{table}[t]
\centering

\begin{tabular}{l|c|c|c|c}
\hline
  & \multicolumn{2}{c|}{Behavioral Approaches} & \multicolumn{2}{c}{Linguistic Approaches} \\
Metric    & Troll Score-based & IRL-based \cite{luceri2020detecting} & Addawood et al. \cite{addawood2019linguistic} & Im et al. \cite{im2020still} \\ \hline
Accuracy  & \textbf{79.5}    & 75.6      & 93.4    & 91.5 \\ \hline
AUC       & \textbf{80.1}    & 81.8      & 98.3    & 96.9 \\ \hline
Precision & \textbf{83.4}    & 76.9      & 92.0    & 90.8 \\ \hline
Recall    & \textbf{81.4}    & 73.4      & 90.0    & 88.1 \\ \hline
F1-Score  & \textbf{81.6}    & 74.8      & 90.9    & 89.2  \\ 
\end{tabular}
\caption{Comparison between behavioral and linguistic approaches for the identification of IO drivers}
\label{comparison2}
\end{table}

Similarly to the previous study case, we select accounts involved in at least ten active and passive online activities. We find 145 IO driver accounts that we aim to classify against a set of 400 randomly selected organic users. We generated trajectories considering L = 100, which lead to a total of 2,479 organic users' trajectories and 900 IO drivers' trajectories.
By following our approach, we perform a classification of the two classes of accounts, whose results are reported in Table \ref{comparison2}. 
Our approach achieves promising performance, with an AUC of 81.8\%, confirming its capability and potential to generalize to different drivers of influence campaigns operating in various countries. 
It is worth noting that, even in this challenging scenario, our proposed methodology surpasses the other behavioral approach \cite{luceri2020detecting} in terms of precision, recall, and F1-Score. While it does not achieve the same classification performance as linguistic approaches, this outcome is promising and underscores the generalization capability of our approach, all while maintaining the characteristic of not requiring access to any content shared in users' messages.

Finally, 
Figure \ref{trollscore2020} displays the Troll Score of the analyzed users and IO drivers. It can be observed that our approach clearly identifies about 70\% of users and IO driver accounts that have a Troll Score of 0 and 1, respectively. However, the remaining fractions of IO drivers and users are more difficult to distinguish, leading to a natural trade-off between true and false positive rates. It is worth noting that our threshold can be adjusted to build a more conservative model and minimize inaccurate classifications of organic users. This is particularly important to reduce social media providers' overhead in the case of misclassification, which may result in moderation intervention such as suspension or ban. Moreover, the different patterns in Figure \ref{trollscore2020} compared to Figure \ref{trollscore200} may reflect a behavioral evolution in IO drivers' deceptive actions, including coordinated strategies and automation \cite{suresh2023tracking,yang2023anatomy,ferrara2023social}.


\section{Conclusion}
In this paper, we introduced a novel two-step AI approach for the detection of troll accounts based solely on behavioral cues. The first step employs an LSTM neural network to classify sequences of online activities as either troll or user activity. In the second step, we utilize the classified sequences to calculate a metric named the ``Troll Score", which quantifies the extent to which an account exhibits troll-like behavior. 
Our initial experiments primarily focused on identifying Russian troll activity on Twitter during the 2016 U.S. election.
The results show that our approach identifies account sequences with an AUC of nearly 99\% and, accordingly, classifies troll and user accounts with an AUC of 91\%, outperforming existing behavioral approaches. Although it does not reach the same level of performance as language-based techniques, our proposed approach does not require access to any content shared in users' messages, making it particularly robust in the era of growing LLM usage for influence campaigns.
When analyzing the Troll Score of the examined accounts, we observed that the majority of trolls exhibit a high Troll Score, indicating that most of their trajectories are classified as troll activity, while most users display a low Troll Score, suggesting that their trajectories are predominantly classified as user activity. However, intriguingly, the results also reveal that a small fraction of trolls have a low Troll Score, implying that their activity closely resembles the behavior of organic users.
To delve deeper into this finding, we conducted a detailed analysis of trolls and users based on their visited state-action pairs. Our analysis reveals the presence of three distinct behavioral clusters among the examined accounts, each populated by both trolls and users. Interestingly, trolls within a specific, albeit less populated, cluster exhibit a relatively lower Troll Score compared to trolls in other clusters, indicating that this particular group of trolls behaves more similarly to organic users than other trolls.

Finally, we tested our approach in a different context with the aim of assessing its generalizability to other scenarios and deceptive, coordinated activities. Specifically, we evaluated the performance of our methodology in identifying accounts involved in diverse information operations during the 2020 U.S. election discussion on Twitter. The results reveal that, while the nature of the drivers of these campaigns might vary, including bots and automation, our methodology effectively identifies their activity and produces promising classification results, which will be further validated in our future endeavors.

\begin{figure}[t] 
	\centering	
	{
		\begin{tikzpicture}
		\begin{axis}[ 
		ylabel near ticks,
            width=0.45\textwidth, 
            height=0.4\textwidth,
            legend style={at={(0.5,0.5)}, anchor=center},
		ylabel={Density},
		xlabel={Troll Score},
		xticklabel style={
			/pgf/number format/fixed,
			/pgf/number format/precision=2		
		},
		xmin=0,
		xmax=1.0,
		xtick={0,0.2, 0.4, 0.6, 0.8, 1},
		ymin=0, 
		ymax=1.0,
		ytick={0,0.2, 0.4, 0.6, 0.8, 1.0},
		grid=both,
        grid style={line width=.1pt, draw=gray!10},
        major grid style={line width=.2pt,draw=gray!50},
		]
		\pgfplotstableread{Results/InfoOps_sample_x_y_users_100.txt}{\comp};
		\addplot [color=blue, ultra thick, mark size=1.3, ] table [x index=1, y index=2] {\comp}; 
		
		\pgfplotstableread{Results/InfoOps_sample_x_y_trolls_100.txt}{\comp};
		\addplot [color=black, ultra thick, mark size=1.3] table [x index=1, y index=2]{\comp};
          
		\legend{Users, IO Drivers} 
		\end{axis}
		\end{tikzpicture}
	}
        
	\caption{Troll Score CDF of users and IO drivers with $L$ = 100}	 
	\label{trollscore2020}
\end{figure}

\begin{backmatter}

\section*{Competing interests}
  The authors declare that they have no competing interests.

\section*{Author's contributions}
FE created the software of this study, conceptualized the LSTM-architecture for classification, and drafted the manuscript.
OA conceived the framework, carried out the analysis, and drafted the manuscript.
LL designed and supervised the study, conceived the framework and its implementation, and finalized the manuscript across the review rounds.
SG designed the study, analyzed the results, and finalized the manuscript.
GN contributed to the software implementation, gathered the dataset, and drafted the manuscript.
EF participated in the analysis of the results and finalized the manuscript. 
IS participated in the design of the study and in the writing of the manuscript.
All authors read and approved the final manuscript.

\section*{List of abbreviations}
\begin{enumerate}
    \item SMNs: Social Media Networks.
    \item IRA: Internet Research Agency.
    \item LSTM: Long short-term memory.
    \item IRL: Inverse Reinforcement Learning.
    \item MDP: Markov Decision Process. 
    \item SVM: Support Vector Machine.
    \item ANN: Artificial Neural Network.
    \item KNN: K-Nearest Neighbors.
    \item CDF: Cumulative Distribution Function.
    \item PCA: Principle Component Analysis.
    \item IO: Information Operation.
\end{enumerate}

\section*{Availability of data and materials}
The code and scripts used to implement the methodological framework detailed below are freely available to the research community. Github repository: \href{https://github.com/FatimaEzzedinee/Exposing-Influence-Campaigns-in-the-Age-of-LLMs-A-Behavioral-Based-AI-Approach}{https://github.com/FatimaEzzedinee/Exposing-Influence-Campaigns-in-the-Age-of-LLMs-A-Behavioral-Based-AI-Approach}. 
In compliance with Twitter's terms of service, we will grant complete access to the tweet IDs that were examined in our study, which can be retrieved using Twitter’s API. It should be noted, however, that tweets from trolls may be inaccessible due to the removal of their accounts and associated tweets from Twitter.
      
\section*{Funding}
The authors gratefully acknowledge support by the Swiss National Science Foundation through the Sinergia project (CRSII5\_209250) \textit{CAll for Regulation support In Social MediA} (CARISMA)  and the SPARK project \textit{Detecting Troll Activity in Online Social Networks (CRSK-2\_195707)}.



\bibliographystyle{bmc-mathphys} 
\bibliography{bmc_article}      
\end{backmatter}

\end{document}